\documentclass{emulateapj}
\usepackage[ansinew]{inputenc}
\usepackage{hyperref}

\def\metersecond{$\rm m\,s^{-1}$}
\def\acceleration{$\rm m\,s^{-2}$}
\def\density{$\rm kg\,m^{-3}$}

\shorttitle{Multiple Impacts of Dusty Agglomerates}
\shortauthors{Kothe et al.}
\slugcomment{Accepted by The Astrophysical Journal}

\begin{document}

\title{The Physics of Protoplanetesimal Dust Agglomerates. V.\\ Multiple Impacts of Dusty Agglomerates at\\Velocities Above the Fragmentation Threshold}
\author{Stefan Kothe, Carsten G{\"u}ttler, and J{\"u}rgen Blum}
\affil{Institut f{\"u}r Geophysik und extraterrestrische Physik, Technische Universit{\"a}t zu Braunschweig,\\Mendelssohnstr. 3, D-38106 Braunschweig, Germany}
\email{s.kothe@tu-bs.de}

\begin{abstract}

In the last years, a number of new experiments have advanced our knowledge on the early growth phases of protoplanatary dust aggregates. Some of these experiments have shown that collisions between porous and compacted agglomerates at velocities above the fragmentation threshold velocity can lead to growth of the compact body, when the porous collision partner fragments upon impact and transfers mass to the compact agglomerate. To obtain a deeper understanding of this potentially important growth process, we performed laboratory and drop tower experiments to study multiple impacts of small, highly porous dust-aggregate projectiles onto sintered dust targets. Projectile and target consisted of 1.5 $\mu$m monodisperse, spherical $\mathrm{SiO_2}$ monomers with volume filling factors of $0.15 \pm 0.01$ (projectiles) and $0.45 \pm 0.05$ (targets), respectively. The fragile projectiles were accelerated by a solenoid magnet and combined with a projectile magazine with which 25 impacts onto the same spot on the target could be performed in vacuum. We measured the mass-accretion efficiency and the volume filling factor for different impact velocities between 1.5 and 6.0 \metersecond. The experiments at the lowest impact speeds were performed in the Bremen drop tower under microgravity conditions to allow partial mass transfer also for the lowest adhesion case. Within this velocity range we found a linear increase of the accretion efficiency with increasing velocity. In the laboratory experiments, the accretion efficiency increases from 0.12 to 0.21 in units of the projectile mass. The recorded images of the impacts showed that the mass transfer from the projectile to the target leads to the growth of a conical structure on the target after less than 100 impacts. From the images we also measured the volume filling factors of the grown structures, which ranged from 0.15 (uncompacted) to 0.40 (significantly compacted) with increasing impact speed. The velocity dependency of the mass-transfer efficiency and the packing density of the resulting aggregates augment our knowledge of the aggregate growth in protoplanetary disks and should be taken into account for future models of protoplanetary dust growth.

\end{abstract}

\keywords{accretion, accretion disks -- methods: laboratory -- planets and satellites: formation -- protoplanetary disks}

\section{Introduction}\label{section_Introduction}
The formation of planets around young stars is a process, which was not only a coincidence in our solar system 4.6 billion years ago but also happened around many other stars in our Milky Way. To this day, nearly 500 extrasolar planets have been observed, but the nature of their formation -- especially the first phase in which dust grows to macroscopic bodies -- is still widely unknown. We are certain that planets form in protoplanetary disks, which are clouds of gas and dust around young stellar objects. Unfortunately, we cannot observe the interior of these opaque disks at sufficiently short wavelengths to see the dust growing so that theoretical and experimental investigations are required to help us understanding the processes leading to the formation of planets.

The interactions between the sub-keplerian gas disk and the embedded dust particles lead to relative velocities between the initially micrometer-sized dust grains so that two dust grains gently colliding will stick to each other \citep{WeidenschillingCuzzi:1993,BlumEtal:2000}. Efficient sticking happens for individual dust grains and small dust aggregates, but to form ever larger bodies, also large aggregates need to grow in mass in mutual collisions. However, many laboratory experiments on dust aggregate collisions in the last years have shown that porous aggregates, e.g. in the mm-size range, do scarcely grow \citep[see review by][]{BlumWurm:2008}. It is within the scope of this series to understand the physics of the collisional interactions of protoplanetesimal dust aggregates and hence to understand the conditions under which they can grow at the first stage of planet formation.

In Paper I \citep{BlumEtal:2006}, we described a mechanism to form macroscopic dust agglomerates, which we consider as relevant to protoplanetary dust. These up to cm-sized monolithic dust aggregates are highly porous, with a volume filling factor of $\phi_0=0.15$ (the volume filling factor is defined as the fraction of the aggregate filled with solid material; hence, the porosity is $1-\phi_0=0.85$). The laboratory samples are 2.5~cm in diameter and can be cut or broken into smaller pieces to perform collision experiments with aggregates of different sizes. \citet[][Paper II]{LangkowskiEtal:2008} used these samples and showed that growth of cm-sized dust aggregates is possible when the size ratio between target and projectile aggregate is sufficiently large and the impact velocity is sufficiently high. Collisions of mm-sized aggregates at low velocities (0.2~\metersecond) were studied in Paper III \citep{WeidlingEtal:2009}. All collisions at this velocity lead to bouncing and compaction of the aggregates, increasing the volume filling factor from $\phi_0=0.15$ up to $\phi=0.36$. Compaction at higher velocities was studied by \citet[][hereafter Paper IV]{GuettlerEtal:2009}, who measured static and dynamical dust-aggregate properties to be implemented into a smooth particle hydrodynamics (SPH) collision model \citep[seealso][]{GeretshauserEtal:2010}. This model was used to reproduce laboratory experiments on the fragmentation of aggregates, and in the present paper we will also give deeper insight into the fragmentation of protoplanetary dust aggregates, in particular focussing on multiple collisions.
Fragmentation is a destructive process which, at a first glance, does not seem to aid in the formation of larger bodies. However, when two dust aggregates collide at a velocity above the fragmentation threshold, which is around 1~\metersecond\ for mm-sized dust aggregates \citep{GuettlerEtal:2010}, both aggregates -- or at least the weaker one -- \emph{do} fragment and the consequences of the fragmentation should be understood. \citet{DullemondDominik:2005} noted that fragmentation can even be helpful in explaining the strong IR excess in the SEDs of T Tauri stars. A continuous cycle of growth and fragmentation could retain sufficiently many small grains to explain the observations while an efficient growth process would not. Another positive effect of fragmentation was discovered by \citet{WurmEtal:2005b} in laboratory experiments, in which they collided cm-sized aggregates with larger targets. Both consisted of irregular SiO$_2$ dust with a grain size of 0.1 to 10~$\mu$m and had an overall volume filling factor of $\phi=0.34$. The velocity was in the range of 6 to 25~\metersecond\ and they observed that, although the projectile fragmented, a significant amount of mass, i.e. up to 0.5 times the projectile mass, was transferred to the target. \citet{TeiserWurm:2009a} showed in experiments that this mass transfer is also efficient in a subsequent collision on the same spot. Multiple collisions in a large number were then studied by \citet{TeiserWurm:2009b}, who collided aggregates of approximately 300~$\mu$m in diameter with several targets at a velocity of 7.7~\metersecond. A stream of particles was falling for 1.5~m and colliding with targets of different sizes on which a dusty crust developed. For flat targets, they found that a conical structure forms; further growth was only limited by the minimum angle with respect to the direction of the incoming particles. Although the sticking process in a single collision could not be observed, it is natural to assume that the growth process is similar to the one described by \citet{WurmEtal:2005b}. All these experiments \citep{WurmEtal:2005b, TeiserWurm:2009a, TeiserWurm:2009b} used setups in which the projectiles were shot downwards, i.e. in the direction of gravity. Therefore, gravity might have supported the accretion of mass to the target. Especially for the experiments by \citet{TeiserWurm:2009b}, the influence of gravity leads to a re-accretion of rebound particles after the impact, which increases the accretion efficiency. The authors compare this with a re-accretion induced by the drag of the surrounding gas in the protoplanetary disk. The small ejecta couple to the gas and are carried back to the target body. This effect was first observed in experiments by \citet{WurmEtal:2001a, WurmEtal:2001b} and afterwards discussed by \citet{SekiyaTakeda:2003, SekiyaTakeda:2005}, and \citet{WurmEtal:2004}. The re-accretion requires a gas flow towards the target surface. High-efficiency re-accretion is in principle only valid as long as the mean free path of the gas molecules is larger than the size of the target body. As the experiments showed re-accreation also for higher pressures (or larger targets, respectively), it has been controversially discussed in how far this process is relevant for the growth of km-sized planetesimals.\\
Even without gas drag, the mass transfer in a fragmenting collision can in principle lead to the formation of larger bodies but the question remains whether the necessary conditions for such an aggregate growth -- a sufficient size ratio between projectile and target aggregate (i.e. the projectile needs to be significantly smaller than the target) and a sufficiently large collision velocity (i.e. the velocity has to be above the fragmentation threshold velocity) -- can be met in a protoplanetary disk. Moreover, the efficiency of the mass-transfer process from projectile to target needs to be high enough to make this growth mode important for the protoplanetary dust evolution.

In a novel approach to unify laboratory and theoretical work, \citet{GuettlerEtal:2010} reviewed 19 laboratory experiments on dust-aggregate collisions and compiled them to the first complete dust-aggregate collision model that makes a prediction for the outcome of any arbitrary dust aggregate collision, including sticking, bouncing, and fragmentation. Although the parameter space is huge and so far only sparsely covered by laboratory experiments, this work is an important step to include all the widespread experimental evidences. To find out, which of the many physical processes identified by \citet{GuettlerEtal:2010} really occur under realistic conditions in a protoplanetary disk, \citet{ZsomEtal:2010a} implemented this collision model into a Monte-Carlo growth model. TheirThe found that protoplanetary dust aggregates grow to a certain size (up to centimeters, depending on the detailed conditions and the nebula model), at which point their further growth is inhibited due to bouncing. The simulations by \citet{ZsomEtal:2010a} were zero-dimensional (0D) in space, valid for one location in the protoplanetary disk, and neglecting global particle motion. The picture changed when \citet{ZsomEtal:2010b} included sedimentation, thus performed a simulation in a vertical column of the disk (1D). Due to turbulent mixing in the vertical direction, the size distribution became wider, which also resulted in larger relative velocities. A larger variety in collision velocities and size ratios also leads to a larger variety of collisional outcomes. In the simulations by \citet{ZsomEtal:2010b}, also the fragmentation with mass transfer \citep[S4 in the notation of][]{GuettlerEtal:2010}, as described by \citet{WurmEtal:2005b}, came into play and can hence be regarded as a potential path to larger protoplanetary bodies.

\begin{figure*}[t]
    \center
    \includegraphics[width=14cm]{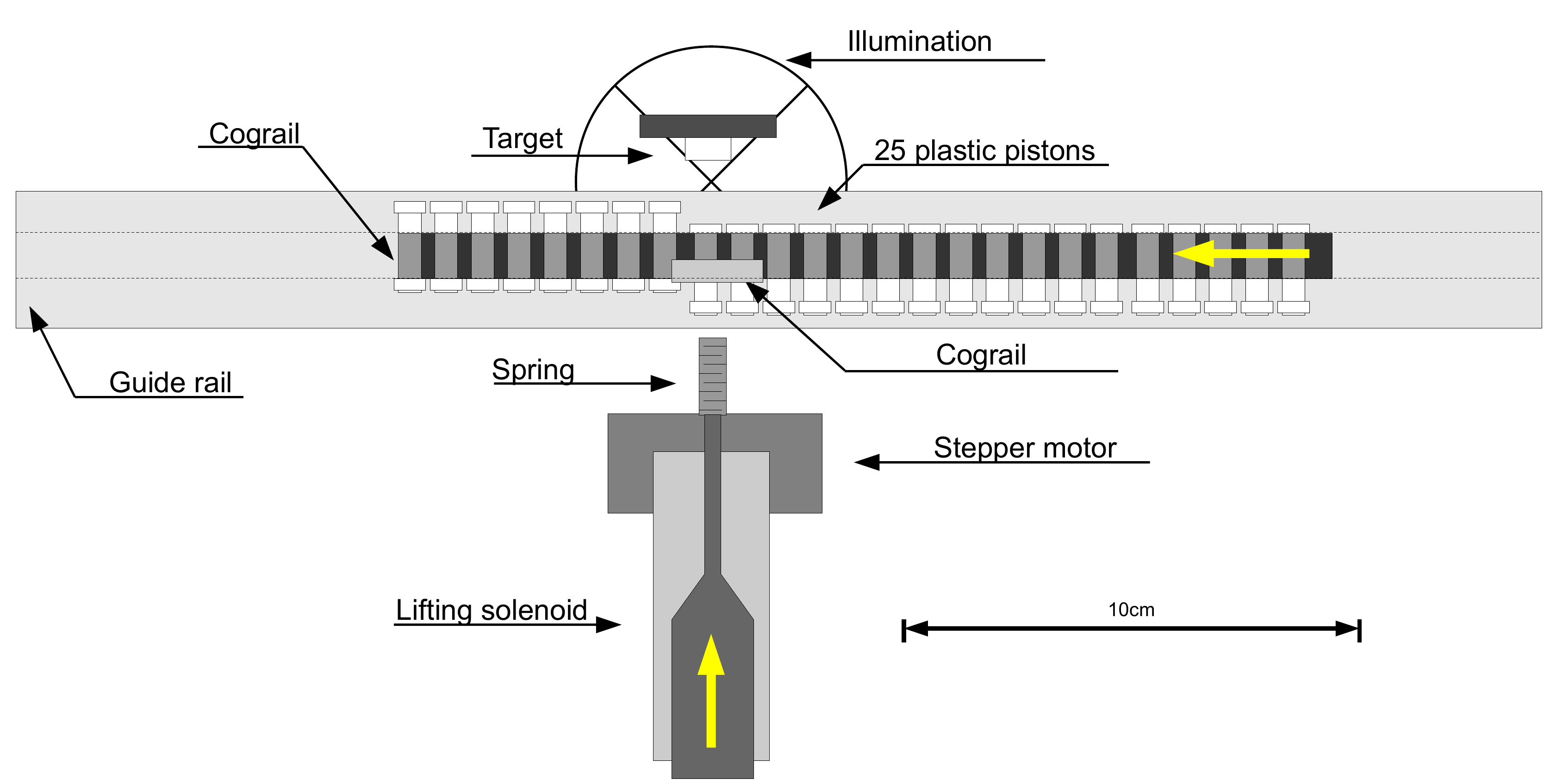}
    \caption{Sketch of the solenoid accelerator and the linear magazine (camera perspective).
    \label{fig.MILA_Skizze}}
\end{figure*}

In this paper, we study the growth of a dust-agglomerate target due to the fragmentation of many mm-sized porous projectiles at velocities between 1.5 and 6.0~\metersecond. This velocity range is chosen to complement previous experiments of \citet{WurmEtal:2005b} and it is moreover in the range predicted by the simulations of \citet{ZsomEtal:2010b}. These high-velocity collisions lead to fragmentation with mass transfer and are slightly above the fragmentation threshold velocity. We study multiple collisions because the accretion efficiency in a single collision is small and so, to get a significant amount of growth, a large number of collisions is required. A self-consistent body forms after the first few impacts have coated the target with a layer of dust. In Sect.\ref{section_Setup}, we will present the experimental setup, experimental procedures, and our dust samples. The results of our laboratory and microgravity experiments are laid out in Sect. \ref{section_Results}. Preliminary results of this work were already implemented in the model by \citet{GuettlerEtal:2010}. While these were very approximate, we will present more details and discuss in Sect. \ref{sec:Discussion} what should be used in protoplanetary dust aggregate collision models in the future. We will also compare our results to those of \citet{WurmEtal:2005b} and \citet{TeiserWurm:2009a, TeiserWurm:2009b}. A conclusion is drawn in Sect. \ref{sec:conclusion}.

\section{Experimental setup}\label{section_Setup}
We developed an experimental setup to perform multiple impacts of dust agglomerates onto a compact dust target. The experiment was designed to shoot up to 25 (mm-sized) dust-aggregate projectiles onto a (cm-sized) dust-aggregate target at velocities between 1.5 and 6 \metersecond\ within five seconds. With this setup, we are capable to study the growth behavior of dust agglomerates in the laboratory as well as under microgravity conditions in the drop tower facility in Bremen (4.7 s free-fall time). The main experiment components are a dust-aggregate accelerator, a fast refill mechanism for the dust aggregates, and a target mounting, which will be explained in detail below (Fig. \ref{fig.MILA_Skizze}). The whole experiment is placed in a vacuum chamber to provide an ambient gas pressure of the order of 0.4 mbar which was available in the drop tower and was also used in the laboratory. To determine the mass gain of the target after one impact series (consisting of up to 25 impacts), we used a laboratory micro-balance.

Although our experiments show some resemblances to the experiments carried out by \citet{WurmEtal:2005b} and \citet{TeiserWurm:2009a, TeiserWurm:2009b}, there are also significant differences. Firstly the impact velocity is smaller than in all experiments before, which we regard as highly relevant. A major difference is also the fact that the projectiles in our setup are shot upwards, i.e. against gravity, or performed in microgravity. With this setup we avoid that the accretion efficiency may be increased due to gravity and we can study the growth process apart from potential re-accretion. Another advantage of the new setup is that a sequence of impacts can be performed within a short time, which enables us to perform a large number of individual, isolated impacts \citep[e.g. compared to the setup of][]{TeiserWurm:2009a}. Still, we can't perform nearly as many impacts as \citet{TeiserWurm:2009b} who also cover the whole surface of the target with impacting projectiles, which is more realistic but is not necessarily advantageous though.

\subsection{Accelerator assembly}\label{subsection_Accelerator}
The impacts in our experiment were realized by a dust-aggregate accelerator, which utilized a 12 V lifting solenoid and a linear dust-aggregate magazine attached to a cograil (see Fig. \ref{fig.MILA_Skizze}). The magazine carries up to 25 (mm-sized) porous dust projectiles, each of which is stored in a dedicated mold on top of a dumbbell-shaped plastic piston. This mounting prevents the projectiles from damage due to the movement through the guiding rail. In each shot, the lifting solenoid moves the plastic piston and, thus, accelerates the projectile, which is then released to a ballistic flight when the piston is abruptly stopped due to its shape (cf. Fig. \ref{fig.MILA_Skizze}). The dust-magazine cograil is continually moving, and to synchronize the shots with the position of the cograil, a photosensor detects the reservoir pistons and triggers the accelerator solenoid. Thus, each experimental series is almost completely automated. The cograil is driven by a stepper motor whose speed is chosen such that the 25 pistons are emptied within less than 4.7 seconds to fulfill the time constrains in the microgravity drop tower. Different impact velocities are possible by adjusting the voltage of the solenoid dust-aggregate accelerator. Due to the inertia of the power supply, the repeatability of the adjusted velocity decreases with increasing voltage. As the porous dust projectiles are very fragile, the acceleration of the solenoid is damped with a spring between plunger and piston.\\
An exchangeable target is placed in a holding mechanism 1 cm above the dust magazine. In the microgravity experiments, this mechanism is connected to a second stepper motor, which turns the target by $180^\circ$ shortly before the drop capsule is decelerated. Once turned, a second lifting solenoid lowers a cap over the target body to prevent mass loss. The impacts are recorded with a high-speed camera with a frame rate of 2,000 Hz and a field of view of $2 \times 1\ \mathrm{cm}^2$. For the back-light illumination of the field of view, we used a 150 W halogen lamp and a diffusor.

\subsection{Dust analog material}\label{subsection_Material}
As analog material for the protoplanetary dust, we used for both, the target and projectile aggregates, monodisperse, spherical SiO$_2$ particles with a diameter of 1.5 $\mu$m. The properties of these grains are well known from measurements of \citet{HeimEtal:1999}, aggregate properties have been measured by \citet{BlumSchraepler:2004} and in Paper I, and it was also used in many of the experiments reviewed by \citet{BlumWurm:2008}. The spherical shape of the dust monomers allows a direct comparison to numerical modeling of dust aggregates \citep[e.g.][]{DominikTielens:1997, WadaEtal:2009} and the same material was also used for the calibration of an SPH code \citep[Paper IV,][]{GeretshauserEtal:2010}.

We used 1.5~mm sized fragments of highly porous dust agglomerates with a volume filling factor of $\phi_0 = 0.15 \pm 0.01$ formed by the method described in Paper I. The fragments were cut out of the larger agglomerate with a razor blade, which causes only a minor compaction at the edge of the agglomerates. Before each experiment, 25 of these fragments were placed in the magazine.

The dust-aggregate target is cylindrical in shape with 1 cm diameter and a thickness of a few mm and is made of compacted and sintered dust. The SiO$_2$ dust particles are compacted with forces from 0.6 to 3~kN, which yield a volume filling factor of about 0.4 to 0.5 (Paper IV). These pellets were then sintered for one hour at $1100^\circ$ C. The sintering prevents the target from damage by the impacts and thus provides an indestructible target from the same material. The surface structure of the target is not changed by the sintering process and is corresponding to a plain layer of dust and allows a self-consistent transition from the indestructible target to the self consistently grown dust layer. Before each experimental sequence, the target was weighted (see below) and placed in the target holding mechanism.

\begin{figure}[tb]
    \center
    \includegraphics[width=6cm]{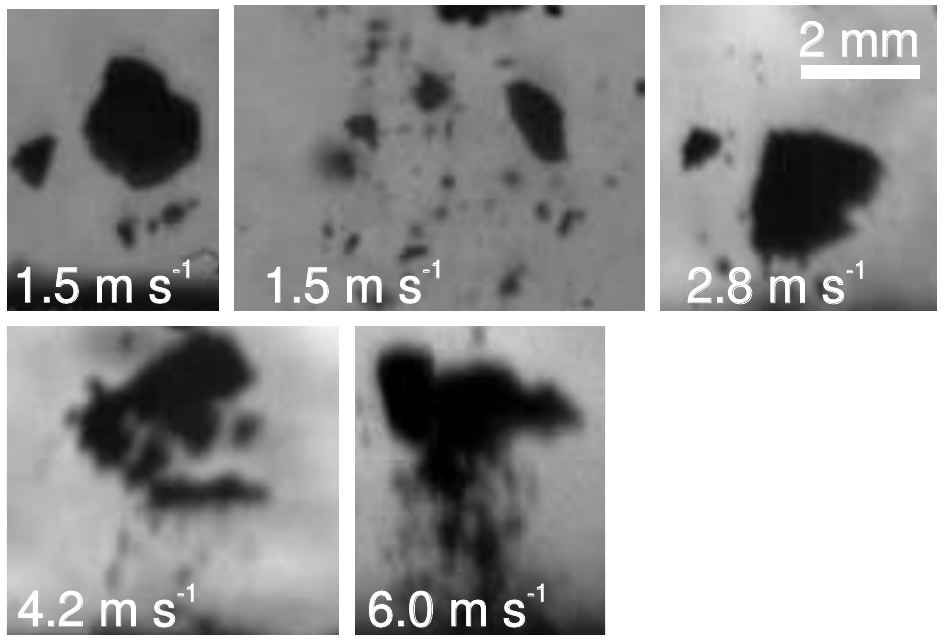}
    \includegraphics[width=\columnwidth]{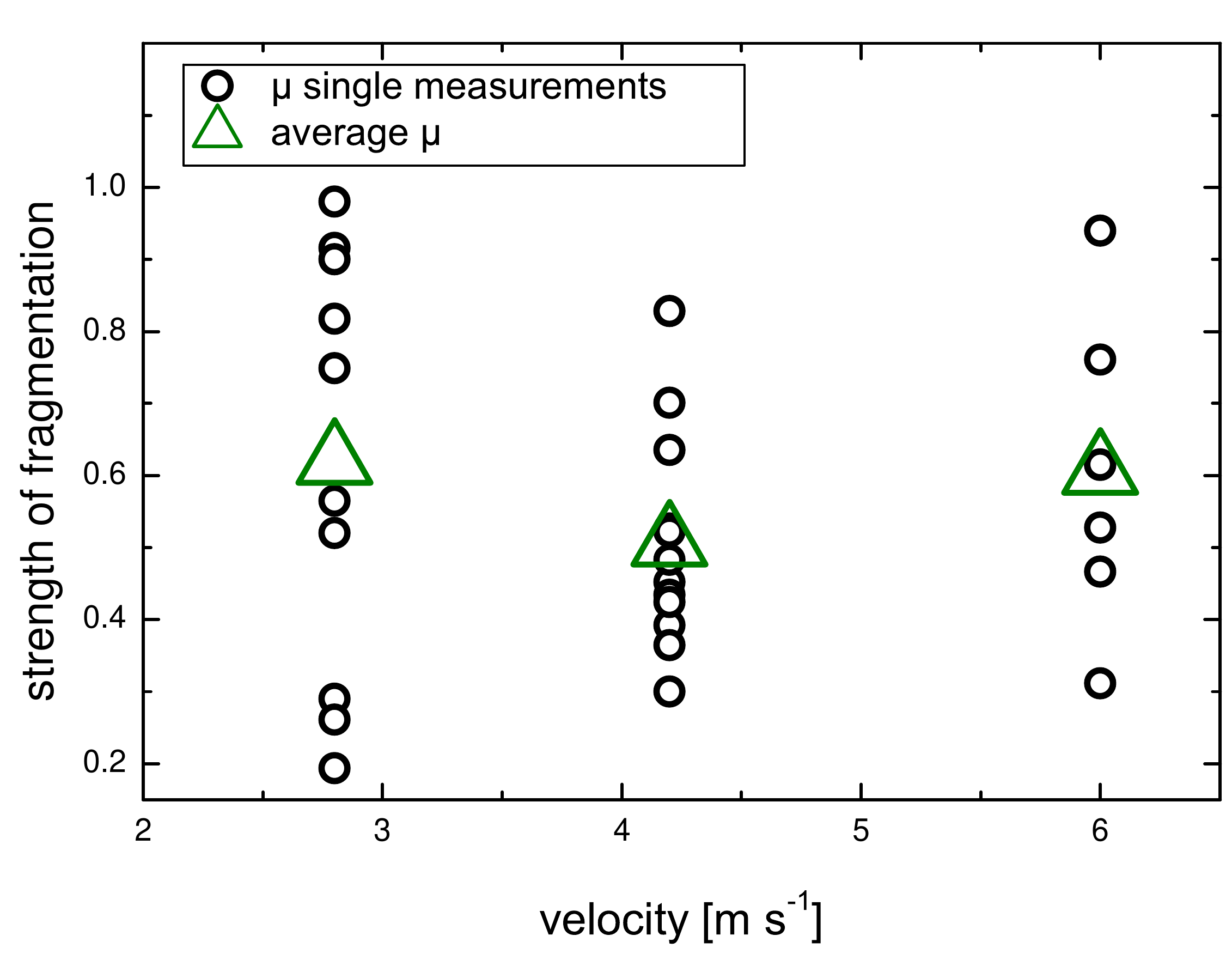}
    \caption{\label{fig.fragmentation}\emph{Top:} Examples for the pre-fragmentation of the projectiles. \emph{Bottom:} The strength of fragmentation (mass fraction of the largest fragment) for different impact velocities.}
\end{figure}

\subsection{Mass determination and projectile properties}\label{subsection_Mass}

The mass of the target was determined after each sequence of 25 shots in the laboratory and after each series (of typically a little less than 25 shots) in the microgravity experiment. Since the dust adsorbs air humidity, the determination of the mass and, thus, the calculation of the volume filling factor and the accretion efficiency must be conducted with great care. First experiments showed that the measured mass of larger amounts of dust fluctuates between different weighings and moreover grows on average with the time after the aeration. To estimate the error caused by this effect, we used the standard deviation of five different weighings within a short time. This error is typically of the order 0.2 mg and was found to be independent of the dust mass in the relevant range. Thus, we regard this error as a limitation to the accuracy of the balance with respect to its place of installation. For the drop tower experiments and early laboratory series (series 1-$\mu$g, 3a-lab, and 3b-lab, cf. Table \ref{tab.exp_overview}), we used this mass determination with a time delay of less than 5 minutes between aeration and weighing. For the later laboratory experiments (series 2-lab and 4-lab), we used an air-tight weighing container to fix the content of water. Continuous measurements over half an hour showed that the mass also grows during the monitored weighing process, which we attribute to (slowly diminishing) surface charges on the glass container. The temporal growth in mass can be approximated by an exponential decay function
\begin{equation}
    M(t) = M_\mathrm{real} - \Delta m e^{-t/\tau} \label{Eq._Mass_det}\ ,
\end{equation}
where $M(t)$ is the measured target mass at time $t$ after the aeration and $\Delta m$ and $\tau$ are fitting constants of the order of 0.5 to 1.5 ~mg and 4 to 12 min, respectively. The value $M_\mathrm{real}$, to which Equation (\ref{Eq._Mass_det}) converges, is our mass determination for which we still use an error of 0.2 mg mentioned above to consider the accuracy of the balance. To reduce the influence of variations in the air humidity (i.e. water content in the container), we furthermore used the differences between two weighings before and after each experimental sequence for further evaluations.
With respect to the mechanical sensitivity of the porous projectiles, it is not feasible to determine the mass of every single projectile. Likewise, it is not possible to weigh the whole magazine due to its large size and mass. We therefore used an average projectile mass determined from 100 representative particles. This value is $m_\mathrm{p} = (1.0 \pm 0.4)$ mg, where 0.4~mg is the standard deviation of the measurements. Here, the influence of the air humidity is measured to be less then 3\% of the mass and can be neglected compared to the large scatter in the individual dust-aggregate masses.

A drawback of the current accelerator is that most of the projectiles broke into a few large pieces due to their rapid acceleration (Fig. \ref{fig.fragmentation}, top). To quantify this breakup, we measured the degree of pre-fragmentation for each velocity regime in at least six shots for each velocity. We shot the projectiles 6~cm high, at which point all fragments were clearly separated, and imaged the size of the largest fragments. The mass fraction of the largest fragment compared with the original projectile is a measure for the strength of the fragmentation, i.e.
\begin{equation}
  \mu = \frac{m_\mathrm{largest\, fragment}}{m_\mathrm{total}} \ .
\end{equation}

The comparison in Fig. \ref{fig.fragmentation} (bottom) shows that the mean fragmentation strength is of the order of $\mu=0.6$ but shows no correlation with the acceleration voltage. The green triangles are the mean values for each velocity regime and do not show a clear trend in velocity. Finally, we investigated whether the projectiles were compacted during the acceleration or due to the vibrations of the experiment. At first, we calculated the compression, which affects the projectile during acceleration. With high-speed imaging of the piston, we measured an acceleration $a$ of the order of 3000, 4000 and 7000~\acceleration\ for the experiments at 2.8, 4.2, and 6.0~\metersecond, respectively. This acceleration corresponds to a pressure of
\begin{equation}
  p = \frac{ma}{A}\ ,
\end{equation}
where $m$ is the mass of the projectile and $A$ is the cross section area of the projectile (i.e. the lower face, which is supported on the piston) of the dust aggregate. For cubic particles with an edge length of 1.5 mm (corresponding to a mass of 1~mg) and a mass density of the dust-aggregates of 300~\density\ (material density of the dust monomers of 2000~\density\ and volume filling factor of 0.15), we thus get pressures of 1.4, 1.8, and 3.2~kPa, respectively. \citet{BlumSchraepler:2004} measured the unidirectional compression of the same material. For the data given by \citet{BlumSchraepler:2004}, an analytic approximation was presented in Paper IV (Equation (9) and Table 1 in Paper IV), which yields a compression of our aggregates to a volume filling factor of 0.17, 0.18, and 0.21, respectively. We must note that this is the compression of the lower edge and the overall compression is much less as the upper edge is not compressed by overlying mass. Moreover, as the compression is far from being static, and forces acting for a duration of the order of 1~ms only, the overall compaction of the dust-aggregate projectiles due to the acceleration process may even be less. We will refer to this estimate later in Sect. \ref{section_Lab_Experiments}.

Moreover, we estimated the influence of vibrations, which might effect the projectiles in their pistons. The sources for these perturbations are the vibrations of the magazine due to the stepper motor and the shock due to each shot. To examine their influence on the projectiles, we filled a magazine with dust aggregates and moved it through the guide rail for 20 cycles. Stereo microscope analysis of representative aggregates before and after the treatment showed no surface damage or compaction. To investigate the shock from an adjacently accelerated aggregate (which is clearly the strongest possible shock, as all other aggregates are hardly affected), we uncovered this piston and used the highest solenoid voltage to push the adjacent piston. We found that the perturbation is sufficiently strong to make the projectile bounce out of the piston. While most of the projectiles were hardly leaving the mold, one out of $\sim 20$ projectiles jumped 13~mm high. This height corresponds to a velocity of 0.5~\metersecond, with which an aggregate would collide with the cover plate and be potentially deformed. However, according to the results of Paper III (Equation 25) this would only cause an additional average compaction of 0.002, which can be considered as negligible.

\section{Results}\label{section_Results}

The following section summarizes the results of the laboratory and the drop tower experiments. For the ground-based experiments, we accomplished four experimental series at three different velocities (2.8, 4.2 and 6.0 \metersecond, respectively), each with 600 single collisions per series and an additional series with 180 impacts at the intermediate velocity. Furthermore, we preformed 11 microgravity experiments at velocities of 1.5 \metersecond\ with a total of 169 collisions. The experiment series are listed in Table \ref{tab.exp_overview}.

\begin{table}[tb]
    \caption{\label{tab.exp_overview}Overview of the performed experiment series.}
    \begin{center}
        \begin{tabular}{ccccc}
          \hline \hline
          series & velocity         & no. of  & accretion  & volume \\
                 & [\metersecond]   & impacts & efficiencies$\mathrm{^a}$            & filling factor       \\
          \hline
          1-$\mu$g & $1.5 \pm 0.6$ & 169 & 0.033$\mathrm{^b}$    & ---$\mathrm{^c}$ \\
          2-lab    & $2.8 \pm 0.2$ & 600 & $0.12 \pm 0.09$ & $0.12 \pm 0.03$ \\
          3a-lab   &               & 600 &                 & $0.24 \pm 0.06$ \\
          3b-lab   & \raisebox{1.5ex}[-1.5ex]{$4.2 \pm 0.3$} & 180 & \raisebox{1.5ex}[-1.5ex]{$0.15 \pm 0.05$} & $0.26 \pm 0.03$ \\
          4-lab    & $6.0 \pm 0.5$ & 600 & $0.21 \pm 0.05$ & $0.40 \pm 0.03$ \\
          \hline
        \end{tabular}
    \end{center}
    $\mathrm{^a}$: mean value of one experiment series of $\sim$ 25 shots.\\
    $\mathrm{^b}$: lower limit for accretion efficiencies.\\
    $\mathrm{^c}$: the volume measurement was not possible as the grown structure was not rotationally symmetric.
\end{table}

First, we will present the accretion efficiency,
\begin{equation}
    e_\mathrm{ac}= \frac{\Delta m_\mathrm{t}}{m_\mathrm{p}} \,
\end{equation}
as calculated from the measured mass growth of the target per shot of one projectile, $\Delta m_\mathrm{t}$, and normalized to one projectile mass, $m_\mathrm{p}$. We will then present the volume filling factor of the accreted mass, which is computed by using the recorded images from the high-speed camera. Section \ref{section_Lab_Experiments} covers the ground-based laboratory experiments, and the microgravity experiments are laid out in Sect. \ref{section_DT_Experiments}.

\begin{figure*}[!tb]
    \center
    \includegraphics[width=\textwidth]{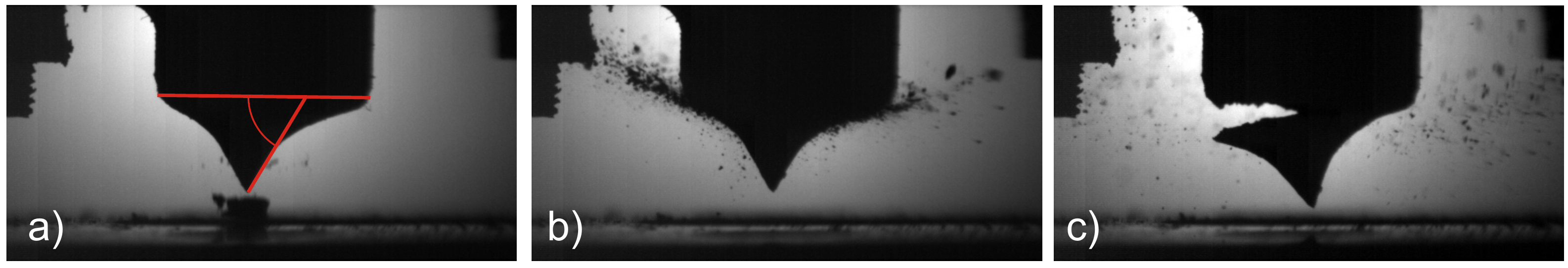}
    \caption{\label{fig.VG12_Abbrechen} \emph{a)} After several impacts with mass transfer (313 in this case), a cone-like structure has formed. A projectile approaching the target is visible right below the tip of the cone. The angle of the cone as described in the text and also used by \citet{TeiserWurm:2009b} is displayed by the red lines. \emph{b)} A typical snapshot of the fragmentation upon impact. \emph{c)} Drop-off of the cone after the impact. The scale of the images is given by the width of the target, which is 1~cm. (An mpeg animation of this figure is available in the online journal.)}
    \includegraphics[width=\textwidth]{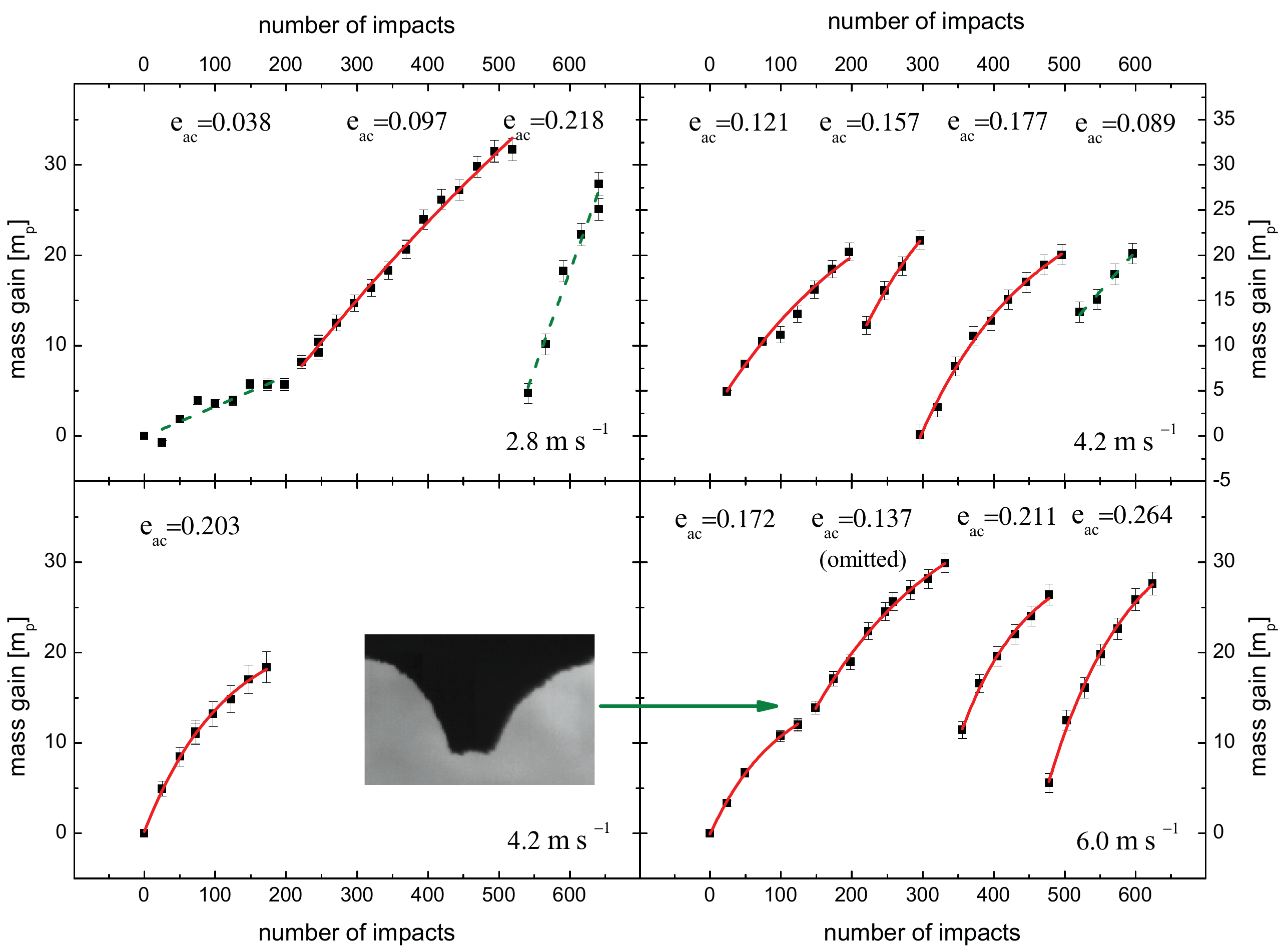}
    \caption{\label{fig.eac_all} Measured mass gain of the target for the laboratory experiments at velocities between 2.8 and 6.0 \metersecond. The data are either approximated by Equation (\ref{Eq. m(n)}) (solid red curves) or by linear functions (dashed green curves). The accretion efficiency calculated from the measurements marked with the green arrow are excluded from later calculations due to partial drop-off of the cone. (Mpeg animations of a series of 25 impacts and a growth sequence, both from experiment 3b-lab (bottom left), are available in the online journal.)}
\end{figure*}

\subsection{Laboratory Experiments}\label{section_Lab_Experiments}

The performed laboratory experiments at velocities of 2.8, 4.2 and 6.0 \metersecond\ showed that repeated impacts lead to the growth of a conical structure at the bottom of the target (Fig. \ref{fig.VG12_Abbrechen}a, b). The growth continues until the structure becomes too heavy and breaks off from the target surface (Fig. \ref{fig.VG12_Abbrechen}c). This happens usually after a few hundred impacts and is often triggered by an impact. To compare our results with those by \citet{TeiserWurm:2009b}, we used the high-speed images to calculate the maximum angles of the cones as illustrated in Fig. \ref{fig.VG12_Abbrechen}a. Ordered by increasing velocities, these angles are $75^\circ$, $59^\circ$, $58^\circ$, and $61^\circ$. It should be noted that due to the breaking off of the cones after a few hundred impacts, these are not necessarily the maximum angles which could have been achieved under microgravity conditions or in top-down collisions as studied by \citet{TeiserWurm:2009a}.

The mass determinations after each series of 25 shots showed a steady increase of the accreted mass on the target (Fig. \ref{fig.eac_all}). After some impacts, this efficiency decreases, probably owing to the increasing impact angle onto the cone. To define the value of the accretion efficiency for impacts on a flat target, or with respect to the collisions of different sized protoplanetary aggregates, it is necessary to calculate the initial slope of accreted mass. This growth can be approximated by an exponentially decaying function of the form
\begin{equation}
   m(N)=m_0 + \delta m \left (1 - \exp \left( \frac{N_0-N}{\xi} \right) \right )\ , \label{Eq. m(n)}
\end{equation}
which we fitted to our data in Fig. \ref{fig.eac_all} (solid red curves). Here, $N$ is the number of impacts, $N_0$ is the first impact number after the cone broke off, $m(N)$ is the measured mass, $m_0$ is the mass at impact number $N_0$, and $\delta m$ and $\xi$ are two free fit parameters. To simplify the matter, we normalize all masses to the mean projectile mass, $m_\mathrm{P}$. The initial accretion efficiency $e_\mathrm{ac}$ onto a flat target is hence given by the slope at impact $N_0$, thus
\begin{equation}
    e_\mathrm{ac}=\left . \frac{\mathrm{d} m(N)}{\mathrm{d} N}  \right |_{N=N_0}= \frac{\delta m}{\xi}\ .
\end{equation}
Some sequences appeared rather linear and could not be fitted by Equation (\ref{Eq. m(n)}). These sequences were then linearly fitted, which is indicated by dashed green lines in Fig. \ref{fig.eac_all}. A linear fit implies a constant accretion efficiency $e_\mathrm{ac}$ which is simply given by the slope of the dashed green lines.

\begin{figure}[!tb]
    \center
    \includegraphics[width=\columnwidth]{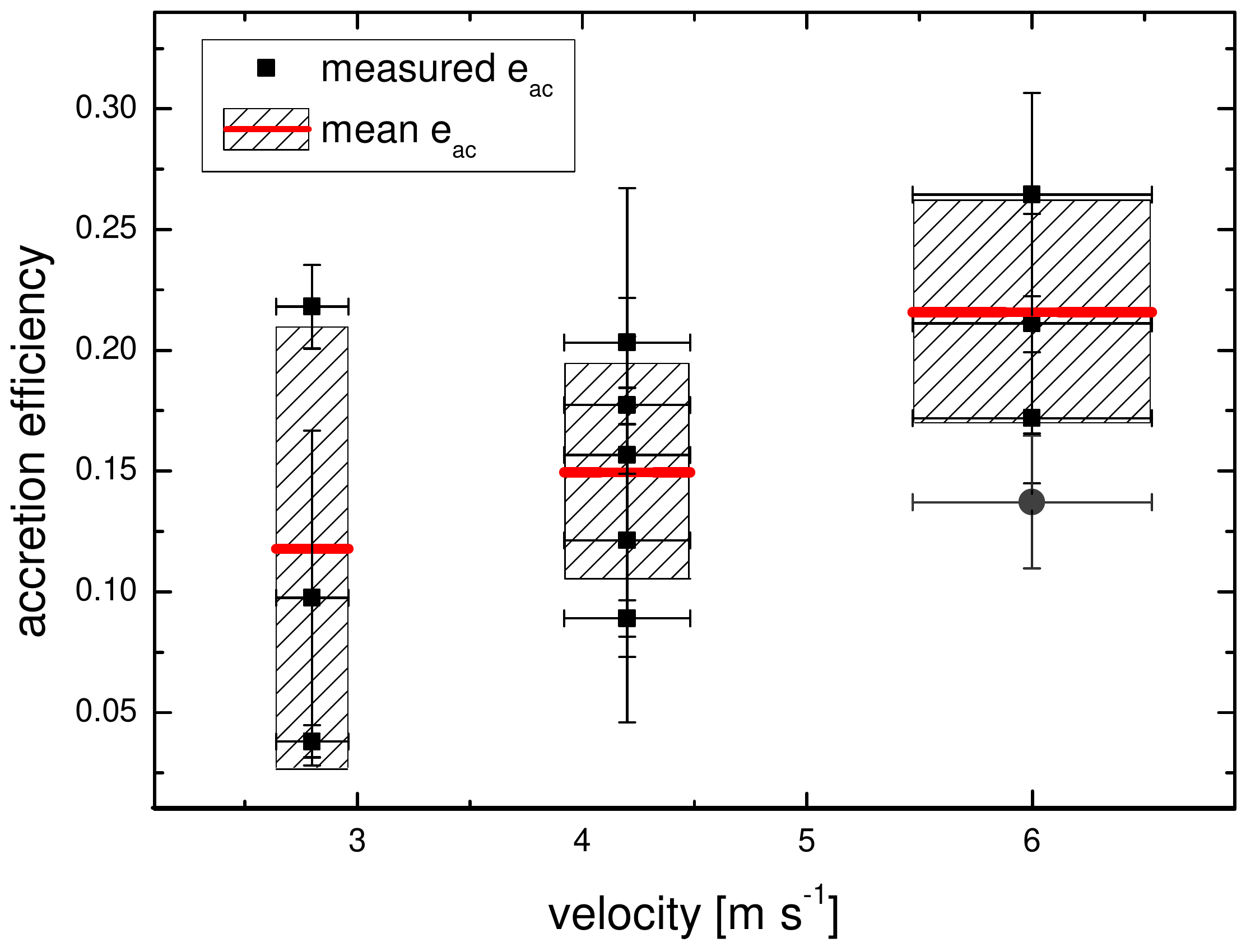}
    \caption{\label{fig.eac}Accretion efficiency, measured in the laboratory experiments, as a function of velocity. The red lines tag the mean values for each velocity range and the boxes denote one standard derivation around the mean values.}
\end{figure}

Figure \ref{fig.eac} shows the derived accretion efficiency as a function of impact velocity. The error of the accretion efficiency follows from the propagation of the errors of the fitting parameters; the error in the impact velocity is the standard deviation of several shots and velocity measurements without target. For each velocity regime, we computed the mean accretion efficiency and the standard deviation, which is indicated by the red lines and the hatched boxes. For the experiments at 6.0 \metersecond\, the value with the lowest accretion efficiency (grey dot) is excluded from the calculation of the mean value, because of the loss of a small part of the cone (cf. inset in Figure \ref{fig.eac_all}). The mean values show a clear trend of increasing accretion efficiencies with increasing velocity. The mean accretion efficiencies and the standard deviation of the single value for each series are given in Table \ref{tab.exp_overview}.

\begin{figure*}[tbp]
    \center
    \includegraphics[width=\textwidth]{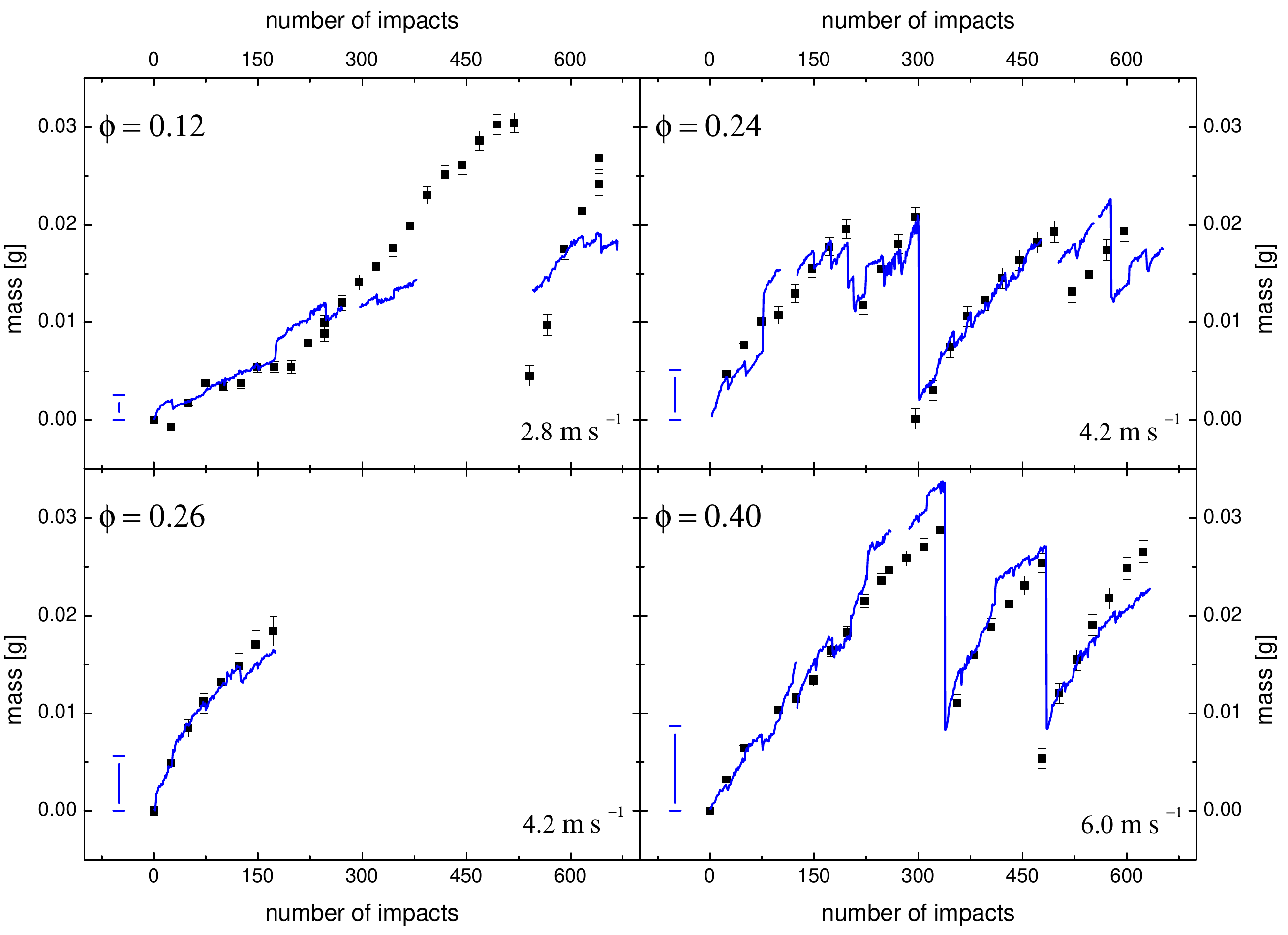}
    \caption{Measured target mass (black squares) and calculated masses from a best fitting volume filling factor $\phi$ (blue lines). The typical error that results from the volume measurements is noted in the left bottom corners. \label{fig.Massen_all}}

\begin{minipage}[hbt]{\columnwidth}
	\centering
    \includegraphics[width=\columnwidth]{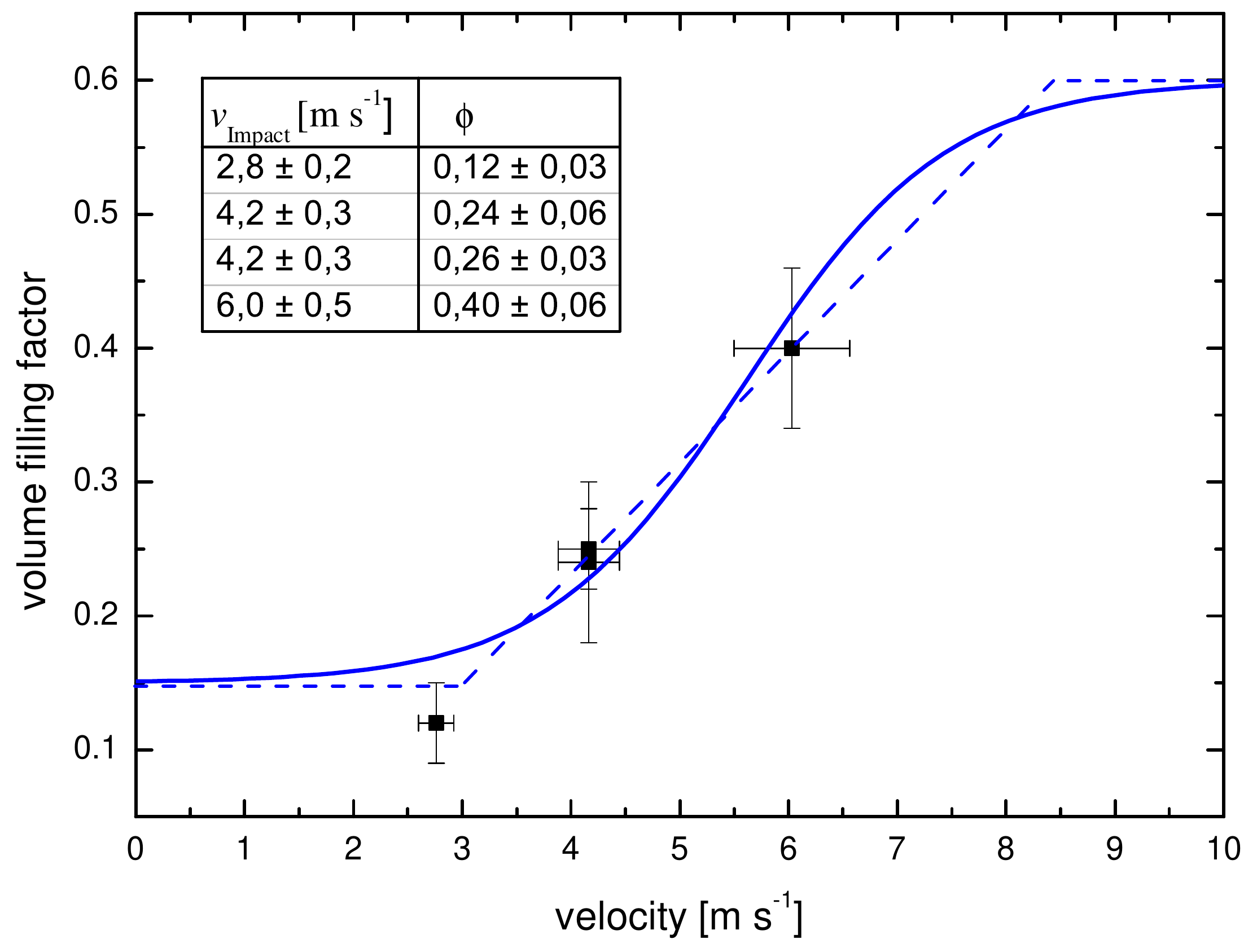}
    \caption{\label{fig.VFF} Measured volume filling factors, from the laboratory experiments, as a function of velocity. The dashed lines denote, from left to right, a constant filling factor of $\phi_0=0.15$, a linear fit to the data, and a constant filling factor of $\phi_1=0.6$. The solid curve corresponds to Equation (\ref{Eq. fermi}).}
\end{minipage}
\hfill
\begin{minipage}[hbt]{\columnwidth}
	\centering
    \includegraphics[width=\columnwidth]{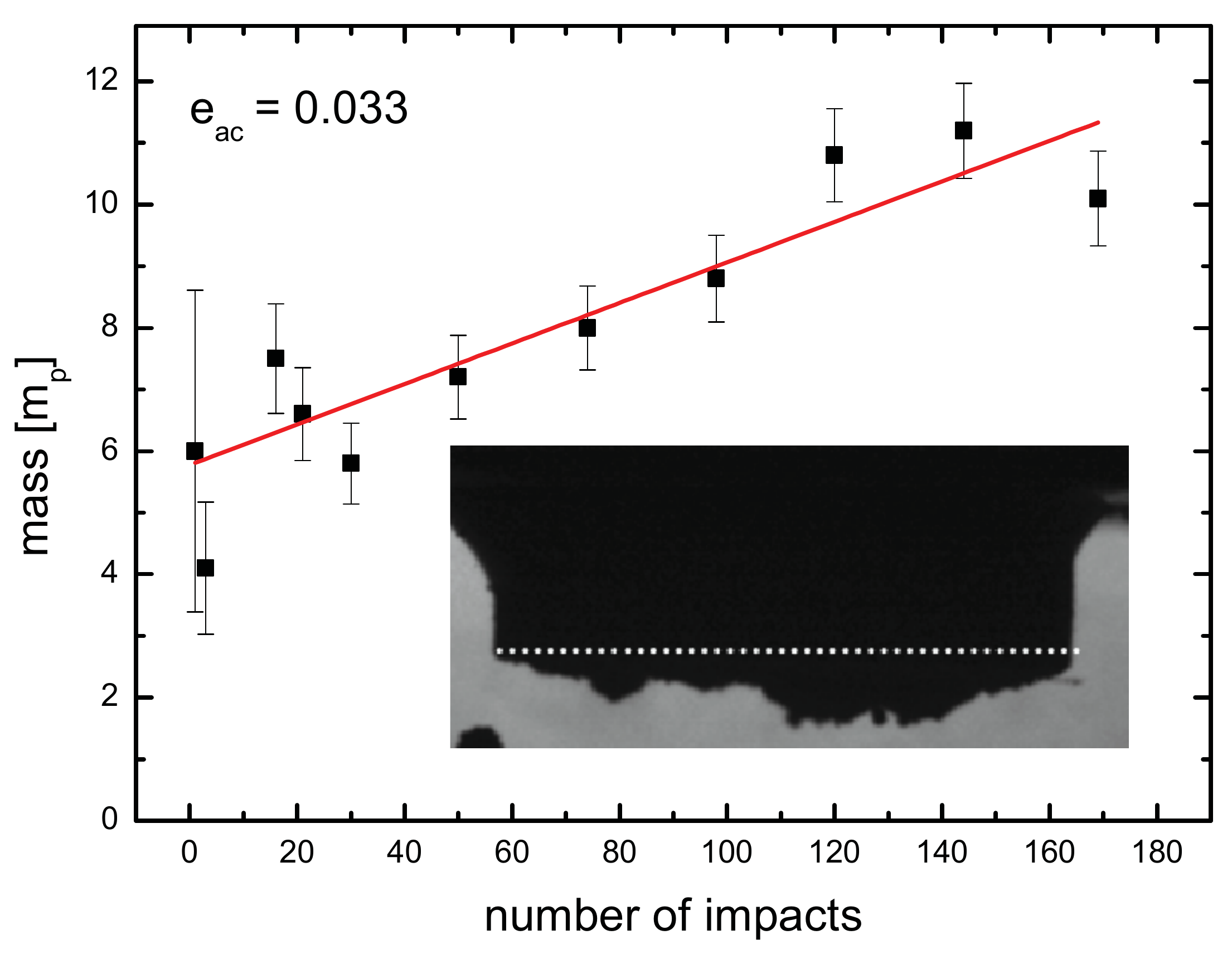}
    \caption{Measured mass gain for the drop tower experiments and a linear fit to the data. \emph{Inset:} Image of the target after 139 impacts. The dotted line marks the position of the target surface.
    \label{fig.waegung_kum}}
\end{minipage}
\end{figure*}

Assuming that the cones have a rotational symmetry, which is supported by our observations, it is possible to estimate their volume filling factors. For a known cone volume $V_\mathrm{cone}$, it can be calculated as
\begin{equation}
    \phi = \frac{M_\mathrm{measured}}{V_\mathrm{cone} \cdot \varrho_\mathrm{grain}}\ , \label{Eq. VFF}
\end{equation}
where $M_\mathrm{measured}$ is the mass of the cone and $\varrho_\mathrm{grain} = 2000$~\density\ is the density of the SiO$_2$ monomers. The volume of the cone can be calculated from the images of the high-speed camera: after each impact the camera image is binarized and the diameter of the cone is measured in each line of pixels. Knowing these diameters, we get a number of disks with a height of 1 pixel, which can be summed up to get the volume of the whole cone. The accuracy of this method is limited by the position, i.e. the height, of the upper edge of the cone. As the cone has its widest extend at this height, a small error in this value leads to a large error in the cone volume. The target was unmounted after each experimental sequence, which leads to a small offset of its position in the images. We therefore correlated a prominent feature of every individual image with a reference image to measure this offset. After correcting this offset in height, the position of the cone edge remains constant over a whole sequence and is determined by the first image in which no cone has grown yet. The remaining fluctuations are of the order of $\pm 2.5$~pixels in height that result from the correlation procedure. They can be treated statistically and so it is possible to calculate the best fitting volume filling factors with their standard derivations. To illustrate the results for the volume filling factor, we used the best value and plotted the mass derived from the images (Fig. \ref{fig.Massen_all}, blue solid line) together with the weighed mass. The stretch factor is the adapted volume filling factor but the shape of the curves reproduces the data very well. For different reasons, no image data was available in some intervals, so the lines in Fig. \ref{fig.Massen_all} can be interrupted. To use image data between two weighings, the mass was linearly interpolated in these intervals. The blue line shows leaps, which are particularly prominent at the impact numbers at which the target was unmounted. These are due to the fluctuations mentioned above, resulting from the detection of the edge. An error of $\pm 2.5$ pixels with a given volume filling factor and target size (i.e. known volume of the upper disks) is indicated by the blue error bars in the lower left corners.

In conclusion, we regard the method described above as sufficiently reliable to compute the volume filling factor and present the results in Fig. \ref{fig.VFF} as a function of velocity (see also Table \ref{tab.exp_overview}). There is an obvious trend for an increasing volume filling factor with increasing velocity. The compaction in the impacts is much stronger than the pre-compaction by the acceleration process, which was discussed in Sect. \ref{section_Setup} and which is henceforth neglected. Our samples had an initial volume filling factor of $\phi_0=0.15$, and lower values are only possible if the cone exhibits macroscopic voids, i.e. when it is made up of the initial aggregates which are intact and loosely bound. Also for the lowest velocity of 2.8~\metersecond, this appeared not to be the case and so we expect that the volume filling factor is rather in the upper range of its estimated error. The idea behind the blue lines will be discussed in Sect. \ref{sec:Discussion}.
\\
\subsection{Drop tower experiments}\label{section_DT_Experiments}
In the additional microgravity experiments at the Bremen drop tower, the impact velocity of $1.5 \pm 0.6$ \metersecond\ was significantly smaller and the degree of fragmentation was clearly lower as in the laboratory experiments described above. While the impact fragmentation in the ground-based experiments was in a nearly fluidized regime and the fragments were small compared to the original projectile (see Fig. \ref{fig.VG12_Abbrechen} b), the collisions at 1.5~\metersecond\ were slightly above the fragmentation threshold velocity \citep[generally assumed to be around 1~\metersecond,][]{GuettlerEtal:2010}. The largest fragments were typically of the order of half the projectile volume. The grown structures in the microgravity experiments were not conical as in the experiments described before but can rather be described as small irregular shaped features, which formed at the lower edge of the target. Although some of these asperities look like small conical structures, they are more likely larger sticking fragments of the damaged projectile. Additionally, due to a mass loss caused during the turn of the target, it was not possible to measure the exact amount of accreted mass after each drop. Observations indicate a mass loss of up to half of the accreted mass. The mass, which remained attached to the target after the target rotation was weighed and is shown in Fig. \ref{fig.waegung_kum}. The mass data were linearly fitted, which yields an accretion efficiency of $e_\mathrm{ac} = 0.033$. Due to the constraints described above, this value can be regarded as a lower limit to the real accretion efficiency. The volume filling factor was computed by a similar method as described above, the only difference is that sticking fragments were individually treated and assumed to be rotational symmetric. This calculation yields a value of $\phi = 0.25$, which is incomparable to the laboratory experiments. Considering the mass loss mentioned above, the measured mass is not overestimated. More likely the volumes of the structures were underestimated due to other sticking dust fragments hiding behind the larger fragments.

\section{Discussion}\label{sec:Discussion}

Our experiments in the laboratory at velocities between 2.8 and 6.0~\metersecond\ are in a similar regime as the experiments by \citet{TeiserWurm:2009b} who used slightly smaller aggregates and an impact velocity of 7.7~\metersecond. In both experiments, the growth of a conical structure was observed, which is only limited by a maximum angle of the slope with respect to the impact direction. Our slop angles, which vary between 58 and 75$^\circ$, are comparable to those of \citet{TeiserWurm:2009b}, who observed angles of 70$^\circ$. One difference of the conical structure is however obvious: while we observed a concave shape, \citet{TeiserWurm:2009b} found a convex structure. A plausible explanation is that in the one case all projectiles hit the same spot while in the other case \citep{TeiserWurm:2009b} the target is exposed to a stream of particles, which hit over the whole surface.

We can also compare the volume filling factor of the cone, which \citet{TeiserWurm:2009b} measured to be $\phi = 0.31 \pm 0.03$ while our maximum value at 6.0~\metersecond\ is already $\phi = 0.40 \pm 0.03$. In Fig. \ref{fig.VFF}, we found a linear relation between the volume filling factor and the velocity. A fit to the data at 4.2 to 6.0~\metersecond\ is shown by the blue dashed line. Values below volume filling factors of $\phi_0=0.15$ and above values of $\phi_1=0.6$ are not physical. The lower value is due to the volume filling factor of the original sample and the higher value describes the maximum value, which can be achieved under high pressures (Paper IV). A smoother empirical approximation of the expected behavior can be given by the approximation function
\begin{equation}
    \phi(v) = \phi_1 - \frac{\phi_1-\phi_0}{\mathrm{exp} \left ( \frac{v-v_\mathrm{m}}{\Delta} \right ) + 1 } \label{Eq. fermi}
\end{equation}
with the two fitting parameters $v_\mathrm{m}=(5.6 \pm 0.5)$~\metersecond\ and $\Delta = 0.9 \pm 0.4$ \metersecond. This corresponding curve is shown by the solid blue line in Fig. \ref{fig.VFF}. Thus, for 7.7~\metersecond, we expect a volume filling factor of the cone of already $\phi = 0.55$, which is well above the value of \citet{TeiserWurm:2009b}. The reason for this might be an additional mass dependency of the volume filling factor. Qualitatively, there can be two explanations for this. We showed in Paper IV (e.g. Fig. 9) that impact compaction is not homogeneous but features a gradient in the volume filling factor, which is due to a peak of the impact pressure at the impact site. Applied to our case, we may assume that the compacted part of the projectile sticks to the target while the fragments can still be very porous. Larger projectile masses will lead to a higher pressure and consequently to a denser packing. Furthermore, the already accreted mass can become further compacted from the following impacts. This compression depends on the mass of the projectile because an impact by a large impactor leads to a higher pressure and thus more compression than that caused by several smaller projectiles with the same total mass. Due to these arguments we also expect a mass dependence of the compaction curve in Fig. \ref{fig.VFF}. However at this point it is not possible to give a more precisely relation from the available data.

\begin{figure}[!tb]
    \center
    \includegraphics[width=\columnwidth]{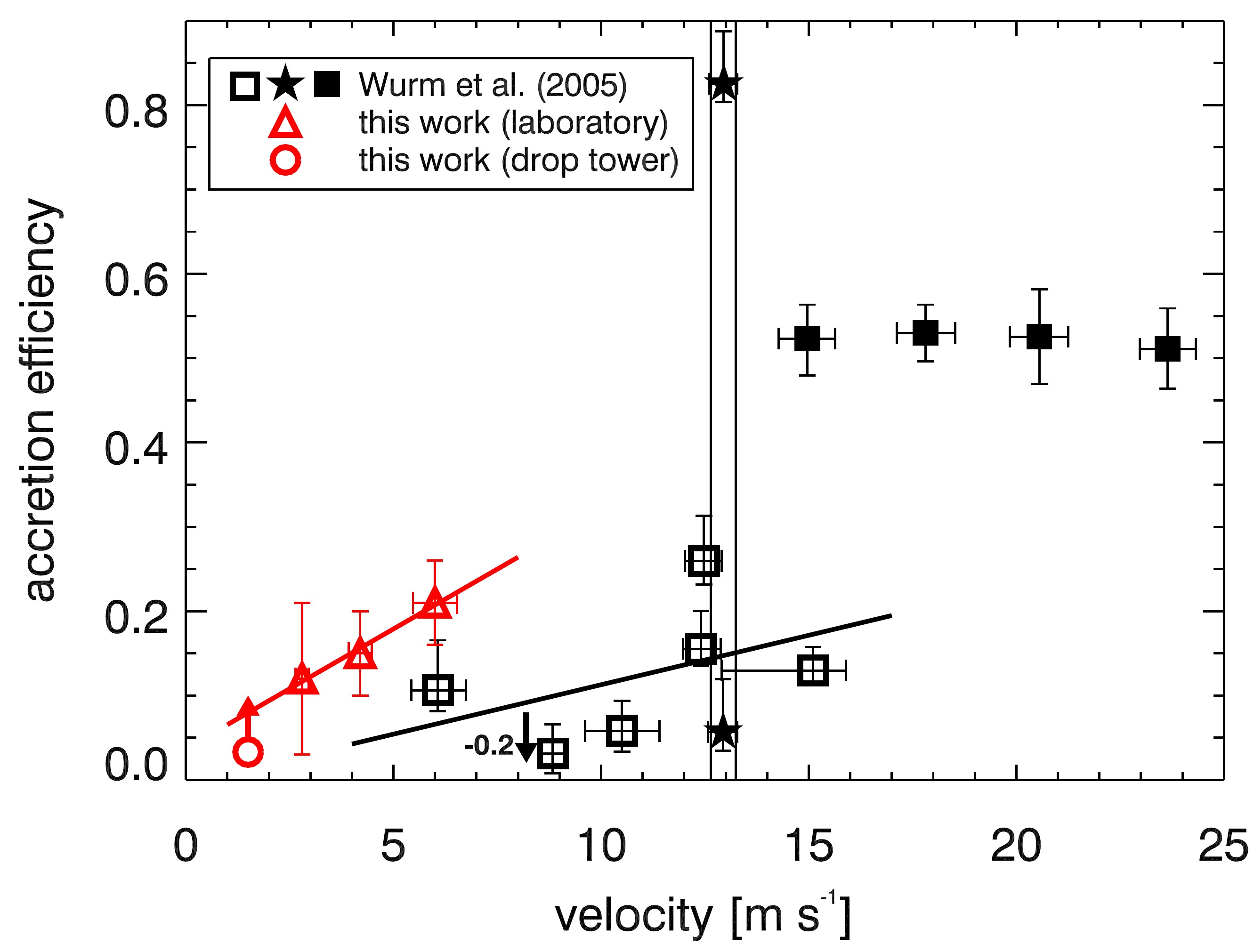}
    \caption{\label{fig:wurm_plot}Accretion efficiency as a function of velocity compared to the results of \citet{WurmEtal:2005b}. The red symbols are the results of this work and the black symbols are those from \citet[][also same notation]{WurmEtal:2005b}. We compare our results to their low-velocity experiments, which are denoted by the open squares.}
\end{figure}

The accretion efficiency as shown in Fig. \ref{fig.eac} seems to follow a linear trend with increasing velocity. This is also in rough agreement with the observations in the microgravity experiments, in which the velocity was just above the fragmentation threshold, although we can only present a lower threshold (see discussion in Sect. \ref{section_DT_Experiments}). Qualitatively, we can again argue with the results from Paper IV (Fig. 7), where we found that the transition between compacted and uncompacted material plays an important role in the strength of the aggregate. We found that the material, which was pulled out of the porous target included the volume which was compacted and broke at the transition to the original volume filling factor. This is due to the transition in the tensile strength, which depends on the volume filling factor as presented in Paper I. Applied to the fragmenting impacts in this paper, we belive that we have a similar gradient in the volume filling factor, i.e. the volume which is adjacent to the target is most compacted. The amount of dust which sticks to the target surface will then depend on the transition between the compressed and uncompressed material. As smaller impact velocities lead to smaller impact pressures, therefore to smaller compacted volumes, the accreted mass and thus the accretion efficiency will also be smaller. We plotted our accretion efficiencies in Fig. \ref{fig:wurm_plot} (red symbols) and applied a fit to the laboratory data (red line). For a linear curve, written as
\begin{equation}
    e_\mathrm{ac}(v) = e_0 + e_1 v \label{eq:accretion_velocity} \,
\end{equation}
we get the two fit parameters $e_0 = 3.7\cdot10^{-2}$ and $e_1 = 2.8\cdot10^{-2}\ \mathrm{m^{-1}\,s}$. In the following, we intend to compare these results with the laboratory experiments of \citet{WurmEtal:2005b}. They used projectiles, which were significantly larger than ours (cylinders with 7.5~mm diameter and 10~mm length), also pre-fragmented, but consisted of a slightly different material (the dust grains were irregular and not monodisperse). Their velocities ranged from 6 to 25~\metersecond, and for the smaller velocities, i.e. 6 to 13~\metersecond, they found an accretion efficiency of the order of $e_\mathrm{ac} = 0.1$. \citet{WurmEtal:2005b} did not focus on this moderate accretion at low velocities, because for higher velocities they found an unexpectedly high accretion efficiency of $e_\mathrm{ac} =0.5$. This effect is stunning but is still not understood, and so we will refer to the experiments in the lower velocity regime, for which an accretion efficiency comparable to our data was derived. We also applied a linear fit to the data of \citet[][open squares]{WurmEtal:2005b}, and we ignored the value with negative accretion, which would not have been possible in our experiments. The respective fit parameters are $e_0 = -4.4\cdot10^{-3}$ and $e_1 = 1.2\cdot10^{-2}\ \mathrm{m^{-1}\,s}$, and the fit is represented by the black line in Fig. \ref{fig:wurm_plot}. The slope of this curve is by a factor of 2.3 lower than for the fit to our measurements. It is likely that this effect is caused by the different projectile masses which vary by a factor of $\sim 100$. This comparison therefore suggests that the accretion efficiency is indeed mass dependent. Beyond that, we cannot make a clear statement if our experiments would reproduce a step in the accretion efficiency like the one observed by \citet{WurmEtal:2005b}, which seems, however, to be unlikely because according to Equation (\ref{eq:accretion_velocity}) an accretion efficiency of 0.5 would already be reached at a velocity of 16.5~\metersecond.

Preliminary results of this work were already included in the model of \citet[][their Sect. 2.2.1]{GuettlerEtal:2010}. The simplified assumptions at that time were that the mass transfer from the projectile to the target is $0.023\, m_\mathrm{p}$ and the volume filling factor is a factor of 1.5 times higher than the volume filling factor of the projectile. The velocity in these experiments was 4~\metersecond, the lower accretion efficiency can be explained by the slightly higher projectile masses (the projectiles in these experiments were intact) and a different target surface. The target was sandpaper and a drop-off of the grown structure occurred earlier for which reason it can be regarded as less stickier. The volume filling factor that went into the model of \citet{GuettlerEtal:2010} was just a best guess based on experiences of other experiments, and a velocity dependence was not expected. In Sect. \ref{sec:conclusion} we will conclude on what should be implemented into future models.

\section{Conclusion}\label{sec:conclusion}
Our experiments confirm that the process of fragmentation with mass transfer is a possible way towards the growth of protoplanetesimal dust aggregates -- always under the condition that in the protoplanetary disk the mass ratio between target and projectile aggregate and the collision velocities are large enough. We thus confirmed the results of \citet{TeiserWurm:2009a, TeiserWurm:2009b} in a way that the accretion efficiency for the first impacts on a flat target remains constantly high, also comparable to the values given by \citet{WurmEtal:2005b}. Future evolution modeling \citep[e.g. in the line of][]{ZsomEtal:2010a} will show whether the size distribution can be wide enough and the velocities high enough to that the accretion process can act. To utilize our results, we propose to keep the following facts in mind:
\begin{itemize}
    \item The accretion efficiency $e_\mathrm{ac}$ is positive and a function of velocity as given by Equation (\ref{eq:accretion_velocity}) (also see Fig. \ref{fig:wurm_plot}, red curve). It also appears to be size dependent as proposed by \citet{TeiserWurm:2009a}, but here we cannot give a quantitative  answer on that.
    \item The accretion efficiency depends on the slope of the grown structure and is maximal for flat targets. This is an artifact of our experiments as similar cones are not expected to appear for protoplanetesimal dust aggregates. However, an obvious implication is that the accretion efficiency will also depend on the impact angle and the surface curvature of the target aggregate.
    \item The volume filling factor $\phi$ of the grown structure is a function of velocity and given by Equation (\ref{Eq. fermi}) (also see Fig. \ref{fig.eac_all}, bottom, blue curve).
\end{itemize}

\acknowledgments \textit{Acknowledgments} We thank the Deutsche Forschungsgemeinschaft for funding this work within the Forschergruppe 759 ``The Formation of Planets: The Critical First Growth Phase'' under grant Bl 298/14-1 and the Deutsche Zentrum für Luft- und Raumfahrt for providing us with drop tower flights.
\bibliographystyle{aa}
\bibliography{literatur}

\end{document}